\documentclass[preprint,12pt,authoryear]{elsarticle}

\usepackage{amssymb}
\usepackage{listings}
\usepackage{url}
\usepackage{geometry}
\usepackage{amsmath}
\newcommand{\og}{\textsc{OpenGadget3}}
\journal{Astronomy and Computing}

\begin{document}
\begin{frontmatter}

%% Title, authors and addresses
%% use the tnoteref command within \title for footnotes;
%% use the tnotetext command for theassociated footnote;
%% use the fnref command within \author or \affiliation for footnotes;
%% use the fntext command for theassociated footnote;
%% use the corref command within \author for corresponding author footnotes;
%% use the cortext command for theassociated footnote;
%% use the ead command for the email address,
%% and the form \ead[url] for the home page:
%% \title{Title\tnoteref{label1}}
%% \tnotetext[label1]{}
%% \author{Name\corref{cor1}\fnref{label2}}
%% \ead{email address}
%% \ead[url]{home page}
%% \fntext[label2]{}
%% \cortext[cor1]{}
%% \affiliation{organization={},
%%            addressline={}, 
%%            city={},
%%            postcode={}, 
%%            state={},
%%            country={}}
%% \fntext[label3]{}

\title{SPACE-Timers - A Stack-Based Hierarchical Timing System for C++}

\author[USM]{Geray S. Karademir}
\author[USM,MPA]{Klaus Dolag}

%% Author affiliation
\affiliation[USM]{organization={University Observatory, Faculty of Physics, Ludwig-Maximilians-Universität München},
        addressline={Scheinerstraße 1},
        postcode={81679}, city={Munich}, country={Germany}}
\affiliation[MPA]{organization={Max-Planck-Institut für Astrophysik},
        addressline={Karl-Schwarzschild-Straße 1},
        postcode={85740}, city={Garching near Munich},country={Germany}}

%% Abstract
\begin{abstract}
SPACE-Timers are a lightweight hierarchical profiling framework for C++ designed for modern high-performance computing (HPC) applications. It uses a stack-based timing model to capture deeply nested execution patterns with minimal overhead, representing runtime behaviour as a tree of timing nodes with precise attribution.

The framework provides structured reports with recursive aggregation, detection of unaccounted time, and compact visual summaries of runtime distribution, supporting both quick inspection and detailed analysis. It also includes checkpointing and error detection mechanisms.

SPACE-Timers supports multiple profiling backends, including NVTX, ITT, ROCtx, and Omnitrace, and integrates with the MERIC runtime system to enable energy-aware optimisation. Its successful use in \og~ demonstrates its effectiveness for large-scale scientific applications.
\end{abstract}

%% Keywords
\begin{keyword}
%% keywords here, in the form: keyword \sep keyword
%% PACS codes here, in the form: \PACS code \sep code
%% MSC codes here, in the form: \MSC code \sep code
%% or \MSC[2008] code \sep code (2000 is the default)
HPC \sep Stack-based timers \sep C++ performance analysis \sep Scientific applications
\end{keyword}

\end{frontmatter}

%% Add \usepackage{lineno} before \begin{document} and uncomment 
%% following line to enable line numbers
%% \linenumbers

\section{Introduction}
\label{sec:intro}
Profiling modern C++ applications with nested execution patterns requires more than flat timing statistics. Developers need hierarchical breakdowns of runtime with minimal overhead instrumentation and integrated structured reporting for performance diagnostics. In contemporary high-performance computing (HPC), where applications often span large numbers of lines of code and execute across heterogeneous architectures, understanding performance bottlenecks at multiple levels of abstraction is essential. Traditional profiling tools frequently either introduce significant overhead or fail to capture the hierarchical structure of modern software, limiting their usefulness in large-scale simulations for daily use.

SPACE-Timers addresses these requirements using a stack-based timing model, allowing developers to measure nested regions naturally. It is particularly suited for HPC simulation codes with timestep loops or multi-phase algorithms (e.g., preprocessing, solving, postprocessing). The project was motivated by the need to replace the manually instrumented timing routines in the \textsc{GADGET} code family \citep{Springel2001, Springel2005} within the context of the upcoming public release of the cosmological N-body/SPH code \og~ (Dolag et al. in prep). During development, it was found that flexible, structured and fine-grained performance insights are essential to systematically improve scalability and efficiency. By combining low-level timing precision with a high-level structural representation of program execution, SPACE-Timers enables developers to systematically analyse runtime behaviour and identify performance-critical regions with minimal intrusion into the codebase. 

\section{Core Concept}
The SPACE-Timers are based on a tree of timing nodes, where each node represents a specific code region with a particular label. Each node accumulated the wall-clock time, and parent-child relationships are reflected in the call hierarchy. This behaviour is provided internally by using the \texttt{std::map} container, with nodes as key-values. Each node stores the accumulated time, the subregions and a pointer to the parent node. The system operates a runtime stack, in which accurate nesting and deterministic attribution of time is ensured. For the actual time measurement \texttt{std::chrono::steady\_clock} is used, which allows high precision timing and cross-platform consistency. The tool relies on the C++17 standard.

\section{Key Features}
The tool provides structured reports featuring recursive time aggregation, sorted output based on each region’s contribution to its parent, and automatic detection of unaccounted time. Unaccounted time is only reported for a region if at least one child region exists and the contribution of all child regions is $\leq99.9\%$ of their parents' runtime. A reduced example of such a report from \og~ is shown in \ref{sec:app1}. These reports support both straightforward inspection and more advanced post-processing, including statistical analysis and cross-run comparisons. In addition, the system provides a symbol-based balance output that compresses the runtime distribution into a concise representation, where each symbol corresponds to a timer region and its length reflects the relative runtime fraction, allowing a quick overview of execution characteristics. Checkpointing capabilities are also included, allowing the full hierarchy stack to be serialised to a file and later reconstructed. Furthermore, the SPACE-Timers incorporate error-handling mechanisms to detect stack underflows, label mismatches, and other potential inconsistencies.

\section{Multi-Backend Instrumentation}
The system integrates with multiple profiling backends, allowing for flexible instrumentation across different hardware and software environments. By default, it supports CPU-based timing, while also providing compatibility with NVIDIA Nsight Systems\footnote{https://developer.nvidia.com/nsight-systems} via NVIDIA Tools Extension Library (NVTX)\footnote{https://github.com/NVIDIA/NVTX}, Intel VTune Profiler\footnote{https://www.intel.com/content/www/us/en/developer/tools/oneapi/vtune-profiler.html} via Intel Instrumentation and Tracing Technology (ITT)\footnote{https://github.com/intel/ittapi} API, as well as ROCTx\footnote{https://rocm.docs.amd.com/projects/rocprofiler-sdk/en/latest/how-to/using-rocprofiler-sdk-roctx.html} developer API and Omnitrace\footnote{https://rocm.github.io/omnitrace/about.html} user API for advanced performance tracing on AMD systems. This modular backend support allows the SPACE-Timers to adapt seamlessly to diverse performance analysis workflows and change between different systems. An example of the simple switch from the base timer to NVTX annotations with NVIDIA Nsight Systems for \og~ is shown in the \ref{sec:app2}.

Additionally, the system includes support for the MERIC\footnote{https://code.it4i.cz/vys0053/meric} energy-efficient runtime system \citep{MERIC}. MERIC is a lightweight C/C++ library designed for HPC applications. It enables dynamic behaviour detection during runtime to reduce energy consumption. Through this integration, the SPACE-Timers extends beyond pure performance profiling to also facilitate energy-aware optimisation strategies in HPC environments.

\section{Usage}
\label{sec:usage}
The SPACE-Timers framework is designed for easy integration into existing C++ applications with minimal code modifications. All interaction with the timers is done through simple macros, allowing users to instrument code regions without directly managing timer objects or internal data structures. Users instantiate a timer hierarchy object for the desired scope and then use macros such as \texttt{TIMER\_PUSH}, \texttt{TIMER\_POP}, and \texttt{TIMER\_POPPUSH} to mark the beginning and end of timed regions. When instrumenting the code is user is forced to provide a measurement level to each push or pop operation. We suggest using $\mathrm{level} = 0$ to instrument only the main function of the code for minimal profiling, by which one only gets the total runtime of the code and to increase the level gradually, e.g. $\mathrm{level} = 1$ for coarse-grained regions, $\mathrm{level} = 2$ as the default balanced profiling and $\mathrm{level} \geq 3$ for fine-grained instrumentation. By setting \texttt{TIMER\_LEVEL} to the level intended at compilation, the user defines the level of instrumentation intended.

Additionally, users can generate structured reports, write symbolic runtime distributions to files, or checkpoint and restore timer hierarchies using macros. This macro-based approach ensures that profiling code is lightweight, non-intrusive, and consistent across different backends and runtime environments. When generating reports, the user can control the depth of the hierarchy (nesting) displayed in the report, omitting any regions that are more deeply nested than the specified level by specifying the public \texttt{Timer\_report\_level} and \texttt{Timer\_balance\_level} variables. To see the effect of changing, e.g. \texttt{Timer\_report\_level} see \ref{sec:app1}.

The cost associated with starting and closing one region (\texttt{TIMER\_PUSH} and \texttt{TIMER\_POP}) is of the order of $\sim10^3$ CPU cycles\footnote{Mean wall-clock time of $\sim2\times10^{-7}$s per execution-pair based on a simple for loop over $10^8$ iterations using an Intel Xeon Gold 6138 CPU at 2.00GHz.}. We have executed several tests with \og, even using a cosmological zoom-in simulation with a base runtime of $\approx60$CPUh. In these experiments, different combinations of \texttt{TIMER\_LEVEL} and \texttt{Timer\_report\_level} were applied to assess the overhead introduced by SPACE-Timers. However, the overhead could not be reliably quantified, as the variation in runtime between repeated runs with identical settings exceeded any differences induced by the parameter changes, and no consistent trend was observed. Given the small number of CPU cycles required, the overhead introduced by the SPACE-Timers in a real application is expected to be negligible.

Since SPACE-Timers are intended for parallel codes, the code offers native MPI \citep{mpi} support and thus can be implemented directly in applications utilising MPI. In the default case, the tool restricts its time measurements to the main MPI rank. This behaviour can be altered by using the \texttt{TIMER\_DETAILS} configuration option, by which the code will then report the time measurements for each rank separately along with basic rank balance statistics including the minimum, maximum, mean, and standard deviation of the runtime across all ranks.

For multi-threaded applications using OpenMP \citep{OpenMP}, timer operations are restricted to the master thread to avoid double-counting and race conditions. Despite its intend to be used with MPI/OpenMP the code also allows serial usage via \texttt{TIMER\_DONT\_USE\_MPI}. For more details on configuration options, see the provided documentation hosted at the code repository.

As an illustration of the output generated by the SPACE-Timers, we present the diagnostic results of a CPU-only blast wave test with \og~ in the panels below. The left panel shows the default reporting configuration (\texttt{Timer\_report\_level} = 2), while the right panel displays the output obtained with an increased reporting level (\texttt{Timer\_report\_level} = 3). The higher reporting level provides more detailed insights into the measured timer regions, while using the same executable at a fixed measurement level (\texttt{TIMER\_LEVEL} = 3). This behaviour is particularly useful, e.g. when restarting a run from a checkpoint in response to performance slowdowns, as increasing the reporting level enables more detailed diagnostics to identify the underlying causes. This small test has a runtime variability of $\leq 0.1s$, which is the reason for the observed faster execution when using a higher reporting level in this particular case. As an example of how the timing levels are assigned in practice, a short pseudocode describing the \texttt{FIND\_HSML} region is provided in \ref{sec:app3}.

\noindent
\begin{minipage}[t]{0.48\textwidth}
\tiny
\begin{verbatim}
Step 100 Time: a=0.07125, MPI-Tasks: 6 Task:0
Total wall clock time for Global = 1.45072 sec
* Timestep                       : 1.4228 sec,  98.07%  
- * FIND_HSML                    : 0.4923 sec,  34.60%  
- - * Secondary                  : 0.2193 sec,  44.54%  
- - * Primary                    : 0.1518 sec,  30.83%  
- - * Exchange                   : 0.0930 sec,  18.90%  
- - * Send_Results               : 0.0144 sec,   2.92%  
- - * Final                      : 0.0096 sec,   1.94%  
- - * Extra                      : 0.0031 sec,   0.64%  
- - * Setup_Left/Right           : 0.0008 sec,   0.16%  
- * HYDRO_ACCEL                  : 0.3354 sec,  23.57%  
...
- * COMPUTE_UNIFIED_GRADIENTS    : 0.2873 sec,  20.19%  
... 
- * check_stop_condition         : 0.1688 sec,  11.86%  
- - * IO                         : 0.0293 sec,  17.35%  
- - * Unaccounted                : 0.1395 sec,  82.65%  
- * DRIFT                        : 0.0392 sec,   2.76%  
...  
- * DOMAIN                       : 0.0167 sec,   1.18%  
...   
- * TREEUPDATE                   : 0.0071 sec,   0.50%  
- * output_log_messages          : 0.0052 sec,   0.37%  
- * SECOND_HALF_KICK             : 0.0029 sec,   0.21%  
- * FIRST_HALF_KICK              : 0.0020 sec,   0.14%  
- * TIMELINE                     : 0.0016 sec,   0.11%  
- * DOMAIN_RECOMPOSITION         : 0.0006 sec,   0.05%  
... 
- * Unaccounted                  : 0.0486 sec,   3.41%
\end{verbatim}
\end{minipage}%
\hfill
\begin{minipage}[t]{0.48\textwidth}
\tiny
\begin{verbatim}
Step 100 Time: a=0.07125, MPI-Tasks: 6 Task:0
Total wall clock time for Global = 1.41166 sec
* Timestep                       : 1.3867 sec,  98.23%  
- * FIND_HSML                    : 0.4790 sec,  34.54%  
- - * Secondary                  : 0.2135 sec,  44.56%  
- - - * HSML_COMPUTE             : 0.1935 sec,  90.67%  
- - - * HSML_WAIT                : 0.0196 sec,   9.17%  
- - - * Unaccounted              : 0.0003 sec,   0.16%  
- - * Primary                    : 0.1500 sec,  31.31%  
- - - * HSML_COMPUTE             : 0.1497 sec,  99.79%  
- - - * Unaccounted              : 0.0003 sec,   0.21%  
- - * Exchange                   : 0.0872 sec,  18.20%  
- - - * HSML_COMM_PREP           : 0.0614 sec,  70.47%  
- - - * HSML_COMM_EXC            : 0.0223 sec,  25.52%  
- - - * HSML_COPY                : 0.0031 sec,   3.54%  
- - - * Unaccounted              : 0.0004 sec,   0.47%  
- - * Send_Results               : 0.0142 sec,   2.96%  
- - - * HSML_COMM_EXC            : 0.0116 sec,  81.70%  
- - - * HSML_COPY                : 0.0023 sec,  16.33%  
- - - * Unaccounted              : 0.0003 sec,   1.97%  
- - * Final                      : 0.0098 sec,   2.04%  
- - - * HSML_FINAL               : 0.0095 sec,  97.53%  
- - - * Unaccounted              : 0.0002 sec,   2.47%  
- - * Extra                      : 0.0031 sec,   0.64%  
- - - * HSML_STATS_EXIT          : 0.0027 sec,  88.12%  
- - - * HSML_UNMARK              : 0.0002 sec,   6.68%  
- - - * Unaccounted              : 0.0002 sec,   5.21%  
- - * Setup_Left/Right           : 0.0010 sec,   0.21%  
- - - * HSML_SETUP               : 0.0009 sec,  83.55%  
- - - * Unaccounted              : 0.0002 sec,  16.45%  
- * HYDRO_ACCEL                  : 0.3242 sec,  23.38%  
...  
- * COMPUTE_UNIFIED_GRADIENTS    : 0.2813 sec,  20.29%  
...
- * check_stop_condition         : 0.1616 sec,  11.65%  
- - * IO                         : 0.0270 sec,  16.72%  
- - - * RESTART_WRITE            : 0.0270 sec,  99.99%  
- - * Unaccounted                : 0.1346 sec,  83.28%  
- * DRIFT                        : 0.0388 sec,   2.80%  
... 
- * DOMAIN                       : 0.0169 sec,   1.22%  
...
- * TREEUPDATE                   : 0.0078 sec,   0.57%  
- * output_log_messages          : 0.0051 sec,   0.37%  
- * SECOND_HALF_KICK             : 0.0030 sec,   0.21%  
- * FIRST_HALF_KICK              : 0.0021 sec,   0.15%  
- * TIMELINE                     : 0.0015 sec,   0.11%  
- * DOMAIN_RECOMPOSITION         : 0.0007 sec,   0.05%  
... 
- * Unaccounted                  : 0.0498 sec,   3.59%   
\end{verbatim}
\end{minipage}

\section{Conclusion}
SPACE-Timers provide a structured, extensible, and efficient solution for hierarchical runtime profiling in C++. Its combination of stack-based timing, rich reporting, serialisation, and backend flexibility makes it suitable for scientific and performance-critical applications. Beyond its core functionality, SPACE-Timers demonstrate that detailed performance insight can be achieved without sacrificing portability or introducing significant runtime overhead. Its integration into large-scale production codes such as \og~ highlights its practical applicability and robustness in real-world HPC environments.

Furthermore, the extensible design enables seamless integration with external profiling and energy-aware runtime systems, thus making SPACE-Timers not only a diagnostic tool but also a foundation for future optimisation workflows. By bridging the gap between fine-grained timing analysis and system-level performance and energy considerations, the framework contributes to the development of more efficient and sustainable scientific software.

\section{License}
Distributed under the \textbf{BSD 3-Clause License}.

\section{Availability}
The source code for SPACE-Timers, including a detailed documentation and small examples, is available at:
\url{https://gitlab.lrz.de/MAGNETICUM/SPACE-Timers}.

\section{General Research Disclosure}
During the preparation of this work, the author(s) used ChatGPT-5 and GPT-5 Mini to obtain suggestions to improve the readability and language of the manuscript, as well as to support documenting and debugging of the software. After using this tool, the author(s) reviewed and edited the content as needed and take full responsibility for the content of the published article and its software.

\section*{Acknowledgements}
We gratefully acknowledge that the initial foundation of the code presented in this release was provided by Martin Reinecke. His work served as the basis upon which this implementation was developed and extended. GSK acknowledges support by the Scalable Parallel Astrophysical Codes for Exascale (SPACE) Centre of Excellence, which received funding from the European High Performance Computing Joint Undertaking (JU) and Belgium, Czech Republic, France, Germany, Greece, Italy, Norway, and Spain under grant agreement No 101093441.

\appendix
\section{Additional diagnostic output}
\label{sec:app1}

To obtain a quick overview of the runtime load balance, the file \texttt{balance.txt} generated with \texttt{Timer\_balance\_report}$= 2$ provides a compact representation of the execution profile. As an example for the same test displayed in Section \ref{sec:usage}, an excerpt for 6 timesteps:
{\tiny
\begin{center}
\begin{verbatim}
  Step=   95 sec=   1.28741     ZZZZZZLJJHHHHHHHGGGSSSSSSSSSSSSSSSSSSSSSSSSSSAAAAAAAAAAAAAAAAPPPPOOIIIIUUUUYY66655433}}''???????????
  Step=   96 sec=   1.21789     ZZZZZZZZZZHHHHHHHGGGGGFFFFFFFFSSSSSSSSSSSSSPPPPIIUUUYY000000066666666554444433]]]]''<<<<????????????
  Step=   97 sec=   1.41703     ZZZLHHHHGGFFFFSSSSSSSSSSSSSSSSSSSSSSSSAAAAAAAAAAAAAAPPPPOOIIIIIUU0000006666666554444}]<<<<??????????
  Step=   98 sec=   1.51453     ZZZJHHHHHGGGGFFFSSSSSSSSSSSSSSSSSSSSSAAAAAPPPPPPIIIIIIUUUUUUUU00000066666655444444}]]'''<<??????????
  Step=   99 sec=   2.15886     ZZZJHHHGGiiiiiiiiiiiiiiiiiiiiiiiiiiiSSSSSSSSSSSSSSSAAAAAAAAAPPPIIIIIUUUUUE000066655443'?????????????
  Step=  100 sec=   1.26858     ZZZZZHHHHGGFFSSSSSSSSSSSSSSSSSSSSSSSSSSSSSAAPPPPPPOIIIIIUUUUUUUUU000000066666655444444433'''????????
\end{verbatim}
\end{center}
}
Each symbol corresponds to a specific timer region, as defined in the following header file. By cross-referencing these symbols with the header, the balance file enables rapid inspection of the runtime composition. This representation is primarily intended for quick, qualitative assessment of the distribution of computational effort across different regions.
{\tiny
\begin{center}
\begin{minipage}{0.5\textwidth}
\begin{verbatim}
Headers for balance file:
'X' - COMPUTE_UNIFIED_GRADIENTS
'Z' - COMPUTE_UNIFIED_GRADIENTS:Exchange
'L' - COMPUTE_UNIFIED_GRADIENTS:Final_Operations
'K' - COMPUTE_UNIFIED_GRADIENTS:Initializing
'J' - COMPUTE_UNIFIED_GRADIENTS:Primary
'H' - COMPUTE_UNIFIED_GRADIENTS:Secondary
'G' - COMPUTE_UNIFIED_GRADIENTS:Send_Results
'o' - DOMAIN
'i' - DOMAIN:DOM_EXCHANGE
'u' - DOMAIN:PEANO
'y' - DOMAIN:RECONSTRUCT_TIMEBINS
't' - DOMAIN:box_wrap
'r' - DOMAIN:check_particles1
'e' - DOMAIN:check_particles2
'w' - DOMAIN:drift_particles
'q' - DOMAIN:find_levels
'M' - DOMAIN_RECOMPOSITION
'N' - DOMAIN_RECOMPOSITION:DOMAIN
'B' - DOMAIN_RECOMPOSITION:PEANO
'V' - DOMAIN_RECOMPOSITION:RECONSTRUCT_TIMEBINS
'F' - DRIFT
'D' - FIND_HSML
'S' - FIND_HSML:Exchange
'A' - FIND_HSML:Extra
'P' - FIND_HSML:Final
'O' - FIND_HSML:Primary
'I' - FIND_HSML:Secondary
'U' - FIND_HSML:Send_Results
'Y' - FIND_HSML:Setup_Left/Right
...
'?' - Unaccounted

\end{verbatim}
\end{minipage}
\end{center}
}
\section{Code integration}
\label{sec:app3}
The following pseudocode shows how the SPACE-Timers are integrated into the \texttt{FIND\_HSML} region in \og, as shown in the diagnostic output in Section \ref{sec:usage}. In this function, the computation of the smoothing lengths (HSML) for particles is implemented. The \texttt{TIMER\_PUSH} and \texttt{TIMER\_POP} macros start and end regions, while \texttt{TIMER\_POPPUSH} transitions between consecutive regions with the same \texttt{TIMER\_LEVEL}. In this example, regions with \texttt{TIMER\_LEVEL} of 1 track fine-grained operations such as computing HSML, preparing communication buffers, and copying particle data, while level 2 timers enclose major phases like setup, primary computation, data exchange, and final calculations. The reason for this choice of level setup in \og~ is that some level 1 regions can occur multiple times at different places; by adding the level 2 regions, one can distinguish between their call origin. This structure enables profiling both individual computation/communication steps and the overall performance of each high-level phase. Although not shown in the pseudocode, \og~ includes one single region with level 0, which is the timestep itself. In addition, one can see how the different \texttt{TIMER\_LEVEL}s differ from the \texttt{Timer\_report\_level}, which is indicated by the indentation in the pseudocode, which can also be observed in the tables in Section \ref{sec:usage}.

{
\begin{center}
\begin{minipage}{0.6\textwidth}
\begin{lstlisting}[language=C++]
function find_hsml()
    TIMER_PUSH(1, "FIND_HSML")        // Start main HSML routine

    // setup phase
    TIMER_PUSH(2, "Setup_Left/Right")
        TIMER_PUSH(1, "HSML_SETUP")
            initialize()  
        TIMER_POP(1, "HSML_SETUP")
    TIMER_POP(2, "Setup_Left/Right")

    iter = 0
    do        // Primary Loop over active particles
        do        // Iterate over all particles
            TIMER_PUSH(2, "Primary")
                TIMER_PUSH(1, "HSML_COMPUTE")
                    compute_primary_hsml()
                TIMER_POP(1, "HSML_COMPUTE")

                TIMER_PUSH(1, "HSML_COMM_PREP")
                    prepare_send_receive_buffers()
                TIMER_POP(1, "HSML_COMM_PREP")

                TIMER_PUSH(1, "HSML_COPY")
                    pack_particle_data_for_export()
                    copy_data()
                TIMER_POPPUSH(1, "HSML_COPY", "HSML_COMM_EXC")
                    exchange_particle_data_with_neighbors()
                TIMER_POP(1, "HSML_COMM_EXC")

                TIMER_PUSH(1, "HSML_COMPUTE")
                    compute_secondary_hsml()
                TIMER_POP(1, "HSML_COMPUTE")

                TIMER_PUSH(1, "HSML_WAIT")
                    MPI_Allreduce_done_flag()
                TIMER_POP(1, "HSML_WAIT")

                TIMER_PUSH(1, "HSML_COMM_EXC")
                    exchange_results_back_to_senders()
                TIMER_POP(1, "HSML_COMM_EXC")

                TIMER_PUSH(1, "HSML_COPY")
                    apply_results_to_local_particles()
                TIMER_POP(1, "HSML_COPY")
        while not all_particles_done()
        
        TIMER_POPPUSH(2, "Primary", "Exchange")
            exchange_particle_info()
        TIMER_POP(2,"Exchange")

        TIMER_PUSH(2, "Final")
            TIMER_PUSH(1, "HSML_FINAL")
                compute_density_and_pressure()
            TIMER_POP(1, "HSML_FINAL")
        TIMER_POP(2, "Final")

    while iteration_needed()        // repeat for particles with insufficient neighbours

    TIMER_POP(1, "FIND_HSML")        // End main FIND_HSML routine
end function
\end{lstlisting}
\end{minipage}
\end{center}
}

\section{Profiler integration}
\label{sec:app2}
The SPACE-Timers framework enables a seamless transition from its internal CPU-based timing stack to external profiling tools, such as NVTX annotations, which can be analysed using NVIDIA Nsight Systems. This functionality is demonstrated using the same test case presented in \ref{sec:app1}.

Before using NVTX annotations with Nsight Systems, the required dependencies must be installed, and the appropriate include and library paths must be configured for compilation and linking on the target system. \og~ is built using \texttt{make}. To enable NVTX-based instrumentation, it is sufficient to add the compilation flag \texttt{-DTIMER\_NVTX} and link against the NVTX library via \texttt{-lnvToolsExt}. No further code modifications are necessary.

\begin{figure}
    \centering
    \includegraphics[width=\textwidth]{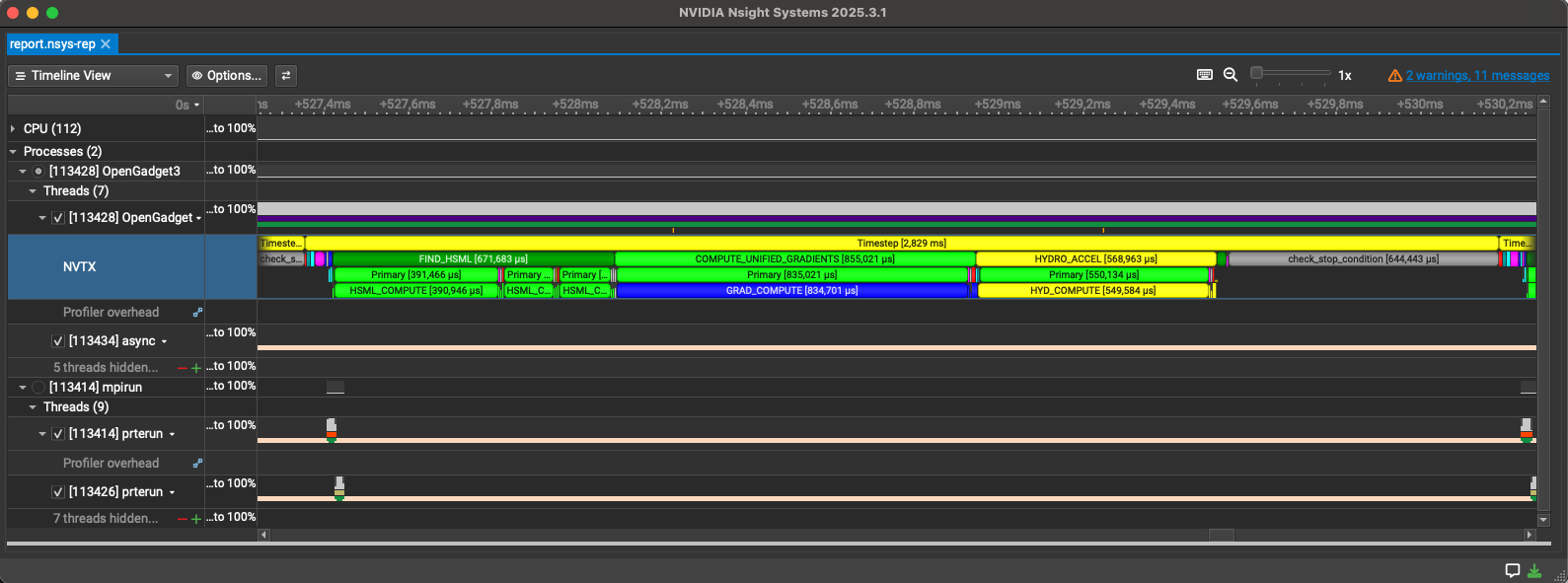}
    \caption{One timestep of the blast wave test run with \og~ visualised in NVIDIA Nsight Systems with NVTX annotation activated. }
    \label{fig:NVTX}
\end{figure}

Enabling NVTX instrumentation in \og~ disables the default CPU timers, resulting in empty \texttt{cpu.txt} and \texttt{balance.txt} output files. All regions previously tracked by the internal timers are instead annotated using NVTX, ensuring equivalent coverage in the external profiler. Running the profile with NVTX support \texttt{nsys profile -t nvtx} will create an \texttt{*.nsys-rep} report file which can be inspected via the NVIDIA Nsight Systems graphical user interface as shown in Fig. \ref{fig:NVTX}. As it can be seen, the same regions as shown in \ref{sec:app1} above are displayed in the timeline view of the profiler, thus allowing for deeper investigation and further code development.

\bibliographystyle{elsarticle-harv} 
\bibliography{sources.bib}

\end{document}